\documentclass[twocolumn,showpacs,prl]{revtex4}

\usepackage{amsmath} 
\usepackage{amssymb} 
\usepackage{graphicx}
\usepackage{psfrag}

\newcommand{\vect}[1]{\boldsymbol{\mathrm{#1}}}

\newcommand{\Z}{\mathbf{Z}}
\newcommand{\C}{\mathbf{C}}
\newcommand{\id}{\mathbf{1}}
\newcommand{\prima}{^\prime}

\newcommand{\sset}[1]{\{#1\}}

\newcommand{\trivial}{\id}

\begin{document}

\title{Topological Order with a Twist: Ising Anyons from an Abelian Model}

\author{H.~Bombin}
\affiliation{Perimeter Institute for Theoretical Physics, 31 Caroline St. N., Waterloo, Ontario N2L 2Y5, Canada}

\begin{abstract}

Anyon models can be symmetric under some permutations of their topological charges. One can then conceive topological defects that, under monodromy, transform anyons according to a symmetry. 
We study the realization of such defects in the toric code model, showing that a process where defects are braided and fused has the same outcome as if they were Ising anyons. These ideas can also be applied in the context of topological codes.

\end{abstract}


\maketitle

\begin{figure}
\psfrag{(a)}{(a)}
\psfrag{(b)}{(b)}
\psfrag{(c)}{(c)}
\psfrag{(d)}{(d)}
\psfrag{a}{\footnotesize$a$}
\psfrag{b}{\footnotesize$b$}
\psfrag{c}{\footnotesize$c$}
\psfrag{j}{\footnotesize $j$}
\psfrag{jp}{\footnotesize $j+1$}
\psfrag{e}{\footnotesize $e$}
\psfrag{m}{\footnotesize $m$}
 \includegraphics[width=8.5cm]{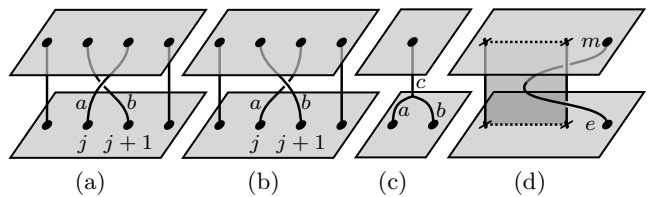}
\caption{Anyon processes: time flows upwards. (a,b) Two topologically distinct ways to exchange anyons. (c) A fusion of two anyons. (d) Two twists (crosses) connected by a line (dotted) across which $e$ charges become $m$ charges.}
\label{fig:braiding}
\end{figure}

Particle statistics are particularly rich in two spatial dimensions, where beyond the usual fermions and bosons there exist more generally \emph{anyons} (see \cite{wilczek:1990:anyons} for a compilation of the basic references). Anyonic statistics are complex enough to give rise to the notion of topological quantum computation (TQC)~\cite{kitaev:2003:ftanyons, freedman:2003:tqc, nayak:2008:anyons}, where computations are carried out by braiding and fusing anyons, see Fig.~\ref{fig:braiding}(a-c). The nonlocal encoding of quantum information on fusion channels and the topological nature of braiding makes TQC naturally robust against local perturbations, providing a complement to fault-tolerant quantum computation~\cite{shor:1996:ftqc, aharonov:1997:ftqc}.

In condensed matter, anyons emerge as excitations in systems that exhibit topological order~\cite{wen:1989:to}. A possible way to obtain these exotic phases is by engineering suitable Hamiltonians on lattice spin systems~\cite{kitaev:2003:ftanyons, kitaev:2006:anyons, wen:2003:toric, levin:2005:stringnet, bombin:2009:2body}. Indeed, implementations on optical lattices have been proposed~\cite{micheli:2006:toolbox}. Unfortunately, the anyon models that appear in simple models are not computationally powerful. In this paper we address an strategy to recover computationally interesting anyon-like behavior from systems with very simple anyonic statistics. 

Our starting point are the symmetries that anyons may exhibit. Anyon models have three main ingredients: (i) a set of labels that identify the superselection sectors or topological charges, (ii) fusion/splitting rules that dictate the charges of composite systems, and (iii) braiding rules that dictate the effect of particle exchanges. A symmetry is a label permutation that leaves braiding and fusion rules unchanged ---for a recent survey, see~\cite{beigi:2010:symmetry}---. Given a symmetry $s$, we can imagine cutting the system along an open curve, as in Fig.~\ref{fig:braiding}(d), and then gluing it again ``up to $s$''. Ideally the location of the cut itself is unphysical, only its endpoints have a measurable effect. In particular, transporting an anyon around one end of the line changes the charge of the anyon according to the action of $s$. Our aim is to explore to which extent these topological defects, that we call twists for short, can be ``treated as anyons'' and used in TQC. Twists are being independently studied by Kong and Kitaev~\cite{kitaev:prep}. An interesting precedent are the Alice strings appearing in some gauge models~\cite{schwarz:1982:electric}, which can cause charge conjugation under monodromy, whereas the twists that we will discuss here exchange electric and magnetic charges.

Rather than trying a general, abstract approach, we will focus on a well-known spin model, the toric code model, and address twists constructively. In this model anyons have no computational power, but we will show that twists behave as Ising anyons~\cite{nayak:1996:ising}, which are computationally interesting. In fact, they do not directly allow universal computation, but there exist strategies to overcome this difficulty~\cite{bravyi:2006:ising, freedman:2006:tilted,bonderson:2010:blueprint}. In~\cite{wootton:2009}, Wootton et al. also try to mimic the non-abelian behavior in an abelian system, using an entirely different approach and philosophy.

We remark that, although the discussion will mainly be in terms of topological order, it has direct application in the closely related context of topological codes~\cite{dennis:2002:tqm, bombin:2006:2dcc, bombin:2007:3dcc, bombin:2010:subsystem}. 

\paragraph*{Anyon models---}

Anyon models are mathematically characterized by modular tensor categories, but we will not need such generalities (for an introduction, see for example~\cite{preskill:2004:notes}). Instead, we will illustrate the content of anyon models with an example: Ising anyons. 

The first element of an anyon model is a set of labels that identify the superselection sectors or  \emph{topological charges} of the model. For Ising anyons there are three: $\trivial$, $\sigma$ and $\psi$. Any given anyon carries such a charge,  which cannot be changed locally. We can also attach a charge to a set of anyons or a to a given region. A region without anyons has trivial charge $1$.
 
Next we need a set of \emph{fusion rules} that specify the possible values of the total charge in a composite system. In terms of anyon processes, fusion rules specify the possible outcomes of the fusion of two anyons, see Fig.~\ref{fig:braiding}(c). For Ising anyons fusion rules take the form
\begin{equation}\label{fusion_Ising}
\sigma\times\sigma=\trivial+\psi, \qquad \sigma\times\psi=\sigma, \qquad \psi\times\psi=\trivial.
\end{equation}
That is, a pair of $\sigma$-s may fuse into the vacuum or produce a $\psi$, a $\sigma$ and a $\psi$ always fuse into $\sigma$ and two $\psi$-s into the vacuum. Fusion rules are commutative and $\trivial\times a=a$.

When two $\sigma$ anyons are far apart, their total charge, which might be $\trivial$ or $\psi$, becomes a non-local degree of freedom. This is indeed an example of a topologically protected qubit, since there are two possible global states. We can measure this qubit in the charge basis by fusing the two $\sigma$-s and checking the output. In general for any set of anyons with given charges there is a fusion space that describes the non-local degrees of freedom related to fusion outcomes. For example, for $2n$ $\sigma$-s with indefinite total charge the fusion space has dimension $2^n$.  

Braiding operations as those in Fig.~\ref{fig:braiding}(a,b) act on the fusion space in a topologically protected way. This action is in general described by \emph{braiding rules}, but in the case of Ising anyons it is possible to characterize braiding with Majorana operators. In particular, for $2n$ $\sigma$-s we need $2n$ majorana operators $c_i$. These are self-adjoint operators that satisfy
$c_jc_k+c_kc_j=2\delta_{jk}$
and act on $\C^{2^n}$, as needed. The total charge of the $j$-th and $j+1$-th anyon is given by the operator $-ic_jc_{j+1}$, the eigenvalues $+1$ and $-1$ corresponding to the total charge $\trivial$ and $\psi$, respectively. Under the braiding of Fig.~\ref{fig:braiding}(a) the operators evolve as follows:
\begin{equation}\label{majorana}
c_j\rightarrow c_{j+1},\qquad c_{j+1}\rightarrow -c_j,\qquad c_k\rightarrow c_k,
\end{equation}
where $j\neq k\neq j+1$. This describes braiding up to process dependent phases that we will not need.

The quantum gates obtained from the braiding of $\sigma$ anyons are not universal. Indeed, they are Clifford gates. Yet, they can be complemented with physically plausible, topologically unprotected, noisy operations to get universal quantum computation~\cite{bravyi:2006:ising}.

\paragraph*{Symmetries---}

We will now illustrate the notion of a symmetry in an anyon model with the model technically known as the quantum double of the group $\Z_2$. It has four charges: $\trivial$, $e$, $m$ and $\epsilon$. The fusion rules are 
\begin{align}\label{fusion_toric}
e\times m&=\epsilon, \qquad e\times\epsilon=m,\qquad m\times \epsilon=e,\nonumber\\
&e\times e=m\times m= \epsilon\times \epsilon=\trivial.
\end{align}
These are \emph{abelian} fusion rules: the result of a fusion has always a definite outcome. Thus, the fusion space is trivial and the effect of braiding a charge $a$ with a charge $b$ as in Fig.~\ref{fig:braiding}(a) can be captured in a phase $R_{ab}$. The $R_{ab}$ phases have in general no measurable effect, but we can build the following invariants, in principle measurable:
\begin{align}\label{braiding_toric}
R_{ee}=R_{mm}&=1, \qquad R_{\epsilon\epsilon}=-1, \nonumber\\
R_{em}R_{me} = R_{e\epsilon}&R_{\epsilon e} = R_{m\epsilon}R_{\epsilon m}=-1.
\end{align}
Thus $e$ and $m$ charges are bosons and $\epsilon$ charges fermions. 

A remarkable property of the rules (\ref{fusion_toric}, \ref{braiding_toric}) is their invariance under the exchange of $e$ and $m$. Indeed, there exists a self-equivalence of the corresponding modular tensor category that exchanges $e$ and $m$ \cite{beigi:2010:symmetry}. It is then conceivable a construction as that in Fig.~\ref{fig:braiding}(d), where there is a line in the system across which $e$ and $m$ charges are exchanged. The rest of this paper is devoted to explicitly realize this in a spin model and explore its consequences.

\paragraph*{Toric code model---}

\begin{figure}
\psfrag{(a)}{(a)}
\psfrag{(b)}{(b)}
\psfrag{x}{\small $x$}
\psfrag{y}{\small $y$}
\psfrag{z}{\small $z$}
 \includegraphics[width=8.5cm]{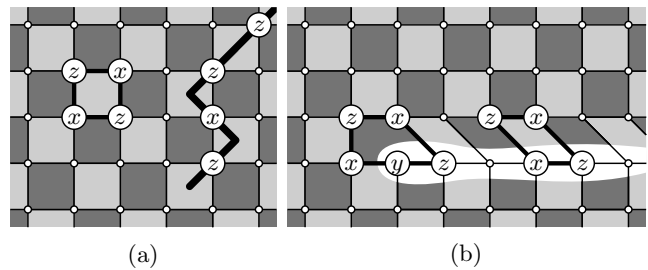}
\caption{A square lattice with spins living at vertices. (a) Plaquette operators are products of four Pauli operators. String operator are products of Pauli operators on their edges. (b) A dislocation in the geometry of the Hamiltonian produced by shifting plaquettes. In the pentagon one can introduce the indicated plaquette operator, which commutes with the rest.}
\label{fig:model}
\end{figure}

This is a spin-$1/2$ Hamiltonian model, with spins forming a square lattice. The interactions are four-body, with each plaquette in the lattice contributing a Pauli product term to the Hamiltonian:
\begin{equation}\label{Hamiltonian}
H:=-\sum_k {A_k},\qquad A_k:= \sigma^x_k \sigma^z_{k+\vect i} \sigma^x_{k+\vect i+\vect j} \sigma^z_{k+\vect j}.
\end{equation}
Here $k=(i,j)$ indexes the spins and $\vect i:=(1,0)$, $\vect j:=(0,1)$. We have chosen a uniform coupling $J=1$; taking instead a different non-zero coupling for each term would not change the physics that interest us.
The Hamiltonian~\eqref{Hamiltonian} might be hard to engineer, but it can be obtained effectively from a two-body spin-1/2 model~\cite{kitaev:2006:anyons} for which there exist experimental proposals~\cite{micheli:2006:toolbox}. The toric code model was originally introduced by Kitaev~\cite{kitaev:2003:ftanyons}, but the more symmetric form~\eqref{Hamiltonian} is due to Wen~\cite{wen:2003:toric}. This symmetry was further studied in~\cite{kou:2008:mutual}.

The ground subspace of~\eqref{Hamiltonian} is described by the conditions $A_k=1$. Excitations are localized and gapped: a plaquette $k$ is excited if $A_k=-1$, in which case we say that it holds a quasiparticle. These quasiparticles are anyons, described by the quantum double of $\Z_2$ (\ref{fusion_toric},\ref{braiding_toric}). To label quasiparticles with their topological charge, we first have to label plaquettes with two `colors' as in a chessboard lattice, see Fig.~\ref{fig:model}(a). Then we can attach a charge $e$ (charge $m$) to quasiparticles living at dark (light) plaquettes. Notice that the exchange of $e$ and $m$ labels is trivially a symmetry at the Hamiltonian level, since the choice of dark/light plaquettes is entirely arbitrary.

\paragraph*{String operators---} 

Consider the action of $\sigma_k^x$. It flips the state of the plaquettes $k-\vect i$ and $k-\vect j$, either hopping a quasiparticle or creating/annihilating a pair of quasiparticles. We can thus visualize $\sigma_k^x$ as a segment $t$ connecting the centers of these two plaquettes and write $S_t:=\sigma_k^x$. Similarly, we relate to a segment $t\prima$ connecting the centers of $k$ and $k-\vect i-\vect j$ the operator $S_{t\prima}:=\sigma_k^z$. Next we consider a string $\gamma$, an ordered collection of segments $(t_i)_1^r$, and build the string operator $S_\gamma:=S_{t_r}\cdots S_{t_1}$. Given an open string $\gamma$, with two different plaquettes $k$ and $k\prima$ as endpoints, $S_\gamma$ flips the eigenvalues of $A_k$ and $A_{k\prima}$. Thus, $S_\gamma$ represents a process where an anyon is moved along the string, plus suitable creation/annihilation events on the endpoints. We can distinguish between dark and light strings, depending on whether they connect light or dark plaquettes; dark (light) strings transport $e$ anyons ($m$ anyons). String operators have two important properties: (i) a dark string $\gamma^D$ and a light string $\gamma^L$ anticommute when they cross and odd number of times, and (ii) if $V_R$ is the subspace of states with trivial charge in a given region $R$ and two strings $\gamma$ and $\gamma\prima$ enclose $R$, then $S_\gamma|_{V_R}=S_{\gamma\prima}|_{V_R}$. That is, strings operators can be freely deformed as long as they do not go over an excitation.

The algebra of string operators alone is enough to compute the anyonic nature of the model~\cite{levin:2003:fermions,kitaev:2003:ftanyons}. Here we are more interested on the fact that among string operators there are constants of motion related to the topological charge of given regions. Notice that closed string operators commute with all plaquette operators; they represent vacuum to vacuum processes. Consider a closed dark string $\gamma^D$ and a closed light string $\gamma^L$ enclosing a given region. Then $S_{\gamma^D}=1$ if the total charge is $\trivial$ or $e$, $S_{\gamma^D}=-1$ otherwise. Similarly, $S_{\gamma^L}=1$ if the total charge in the region is $\trivial$ or $m$, $S_{\gamma^D}=-1$ otherwise. Together, $S_{\gamma^D}$ and $S_{\gamma^L}$ distinguish the four charges.

\paragraph*{Twists---} 

To construct a line $L$ across which $e$ and $m$ are exchanged, as in Fig.~\ref{fig:braiding}(c), we have to introduce dislocations in the lattice: along $L$, plaquettes are shifted so that the coloring does not match, as in Fig.~\ref{fig:model}(b). The geometrical change implies a change in the Hamiltonian, and we indicate in the figure the new plaquette operators. At the ends of the line we find pentagonal plaquettes. We can either add a pentagonal plaquette operator, as indicated in the figure, or allow a localized gapless mode. Turning points of $L$ can be dealt with similarly.

The topologically interesting behavior occurs at the ends of $L$, the twists. A string that winds once around a twist cannot close, as it changes from dark to light or vice versa. That is, $e$ charges become $m$ charges if they are transported around a twist. Because of these global properties, it is clear that twists have a topological nature, in the sense that isolated twists cannot be created or destroyed locally. In particular, the parity of the number of twists in a region cannot be changed without altering the geometry in the boundary of the region.

With twists, the labeling of $e$ and $m$ charges becomes inconsistent. But we can still fix a local labeling at a basepoint and make it global  via a fixed set of paths. Also, from a topological perspective only the location of the twists matters. In particular, if plaquettes are shifted along a circle enclosing a region without twists the topological properties of the system are unchanged.

The possibility of considering dislocated lattices for~\eqref{Hamiltonian} was already pointed out by Kitaev in~\cite{kitaev:2006:anyons}, where he argued the appearance of topological degrees of freedom. We will see not only that this is true, but also that twists have interesting fusion and braiding rules.

\paragraph*{Generalized charges---}

\begin{figure}
\psfrag{(a)}{(a)}
\psfrag{(b)}{(b)}
\psfrag{(c)}{(c)}
\psfrag{(d)}{(d)}
\psfrag{(e)}{(e)}
\psfrag{s}{$=$}
\psfrag{sm}{ $=-$}
\psfrag{dots}{$\dots$}
 \includegraphics[width=8.5cm]{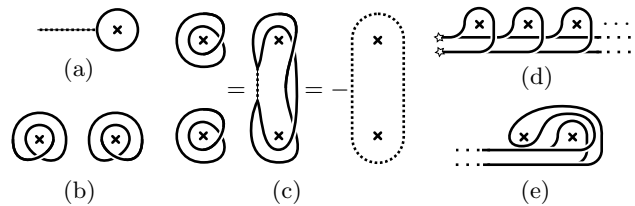}
\caption{Several string operators (solid lines) and twists (crosses), in a top view of the system. The ordering of segments in the string dictates which is drawn over which. Dotted lines are fermion string operators, two parallel string operators of different type. 
(a) A process that creates/annihilates an $\epsilon$. (b) Two inequivalent closed string operators winding around a given twist. We choose the one to the left to define $\sigma_+$ and $\sigma_-$. (c) Three equivalent string operators. The first equation uses property (ii) of string operators, the second uses also (i). (d) The strings $\gamma^1, \gamma^2, \gamma^3, \dots$ that define Majorana operators. The stars mark their endpoints. (e) The effect on $\gamma^j$ strings of braiding adjacent twists as in Fig.~\ref{fig:braiding}(a) .}
\label{fig:strings}
\end{figure}

Twists act as sources and sinks for $\epsilon$ fermions. Indeed, an $\epsilon$ fermion can split in a pair of $e$ and $m$ charges, and if one of them winds around a twist they can annihilate, see Fig.~\ref{fig:strings}(a). This suggests that we cannot distinguish four topological charges in a region with a twist, as $\trivial$ and $\epsilon$, and also $e$ and $m$, become identified. String operators make this more precise. In a region with an odd number of twists we cannot define the two enclosing strings needed to distinguish the four charges. Instead, there is a single string $\gamma$ that winds twice around the region, see Fig.~\ref{fig:strings}(b). Since $\gamma$ self-crosses, the ordering of its segments is relevant and we choose one of two topologically distinct orientations, see the figure. The eigenvalues of $S_{\gamma}$ distinguish two new topological charges: $\sigma_+$ for $S_\gamma=i$ and $\sigma_-$ for $S_\gamma=-i$. The pentagonal plaquette operator of Fig.~\ref{fig:model}(b) introduces an energy difference between them. 

We have six generalized topological charges in the model: $\trivial$, $e$, $m$, $\epsilon$, $\sigma_+$ and $\sigma_-$. These are not the charges of an anyon model, because anyonic charges are not modified through braiding. Yet, they do have well defined fusion rules that can be recovered using string operators:
\begin{align}\label{fusion_generalized}
\sigma_\pm\times \sigma_\pm&=1+\epsilon, \qquad \sigma_\pm\times \sigma_\mp=e+m, \nonumber\\
\sigma_\pm\times \epsilon=&\sigma_\pm,\qquad\sigma_\pm\times e=\sigma_\pm\times m= \sigma_\mp.
\end{align}
To exemplify the computation of these rules, consider the fusion $\sigma_+\times\sigma_+$. Let $\gamma^1$, $\gamma^2$ be the strings that wind twice around each of the twists, as in Fig.~\ref{fig:strings}(c), and let $\gamma^L$, $\gamma^D$ be two nonequivalent strings that enclose the two twists. Then, as demonstrated in the figure, in the absence of other excitations $S_{\gamma^D}S_{\gamma^L}=-S_{\gamma^1}S_{\gamma^2}=1$, which implies that the total charge can only be $\trivial$ or $\epsilon$. Since we can switch the charge by moving an external $\epsilon$ ``into'' any of the two twists, the rule follows.

\paragraph*{Ising anyons---} 

In view of~\eqref{fusion_generalized}, it is tempting to consider the subset of charges $\sset{\trivial,\sigma_+,\epsilon}$: it is closed under fusion ---with rules~\eqref{fusion_Ising} via $\sigma_+\rightarrow \sigma$, $\epsilon\rightarrow\psi$--- and a charge $\epsilon$ remains the same after being transported around $\sigma_+$. Could this subset of charges be regarded as an anyon model? Not really: we cannot attach an invariant meaning to $R_{\epsilon\sigma_+}R_{\sigma_+\epsilon}$, because we cannot directly compare~\cite{levin:2003:fermions} the path of an $\epsilon$ that surrounds a $\sigma_+$ and one that surrounds a trivial charge: the geometry is different. Nevertheless, as we will show now, the braiding and fusion of $\sigma_+$ twists reproduces that of $\sigma$ Ising anyons. 

If braiding twists  is to make sense, we need to move them. For example, by adiabatically transforming the ``geometry'' of the Hamiltonian. In a different context, that of toric codes~\cite{kitaev:2003:ftanyons,dennis:2002:tqm}, there is no Hamiltonian but just an encoding subspace that corresponds to the ground subspace of~\eqref{Hamiltonian}. There we can also change the geometry through code deformations~\cite{dennis:2002:tqm, raussendorf:2007:deformation, bombin:2009:deformation}. Although the geometry of the Hamiltonian or code depends necessarily to a certain extent on the braiding history, at a topological level we can ignore this as long as different processes are not compared. Then, whatever the setting, with or without a Hamiltonian, we can analyze the effect of braiding following the evolution of closed string operators, which are topologically protected. We will focus on the braiding of $\sigma_+$ twists, but the treatment of $\sigma_+$ and $\sigma_-$ twists together is very similar.

The key to understand braiding is to introduce strings $\gamma^j$ such that the operators $c_j:=S_{\gamma^j}$ realize the Majorana operators in~\eqref{majorana}. A suitable choice for the $\gamma^j$ is given in Fig.~\ref{fig:strings}(d). All these strings have the same endpoints, so that any product of an even number of $c_j$-s corresponds to a closed string configuration and gives a constant of motion. This is true in particular for the operators $C_j:=-ic_jc_{j+1}$, which give the total charge of the $j$-th and $(j+1)$-th twists. Namely, $C_j=1$ if the charge is $\trivial$ and $C_j=-1$ if it is $\epsilon$, exactly as desired. Moreover, under braiding the evolution of the constants of motion is dictated by~\eqref{majorana}, as can be computed graphically starting from Fig.~\eqref{fig:strings}(e). In conclusion, $\sigma_+$ twists exactly mimic Ising anyons under braiding and fusion. That is, as long as process dependent phases are not involved, which fails for example in an interference experiment.

\paragraph*{Discussion---}  Beyond the particular example addressed here, it would be interesting to explore twists in general anyon models. E.g., can twist braiding and fusion be computationally universal in a non-universal anyon model? Also, we have adopted an approach where twists are directly engineered, but they could appear randomly in systems where topological order emerges naturally. In the context of topological codes, the ideas presented here are tools to design and manipulate codes. E.g., they allow to get planar versions of the topological subsystem codes introduced in~\cite{bombin:2010:subsystem}.

\begin{acknowledgments} 


I am grateful to Janet Hung, Lucy Zhang, Carlos Tamarit, Daniel Gottesman and Jaume Gomis for useful discussions, and very specially to John Preskill and Xiao-Gang Wen for comments on an earlier version of the manuscript. In particular, J.P. pointed me out the work~\cite{kitaev:prep}. This work was supported with research grants QUITEMAD S2009-ESP-
1594, FIS2009-10061 and UCM-BS/910758.

\end{acknowledgments}

\bibliography{refs,comments}

\end{document}